\documentclass[fleqn,usenatbib]{mnras}

\usepackage{newtxtext,newtxmath}

\usepackage[T1]{fontenc}

\DeclareRobustCommand{\VAN}[3]{#2}
\let\VANthebibliography\thebibliography
\def\thebibliography{\DeclareRobustCommand{\VAN}[3]{##3}\VANthebibliography}

\usepackage{graphicx}	
\usepackage{amsmath}	
\usepackage{multirow}

\newcommand{\tr}{\intercal}
\newcommand{\M}{\mathbfss{M}}
\newcommand{\C}{\mathbfss{C}}
\newcommand{\D}{\mathbfss{D}}

\title[Quasi-periodic \& steep-spectrum pulsar noise]{Impact of quasi-periodic and steep-spectrum timing noise on the measurement of pulsar timing parameters}

\author[M.~J.~Keith and I.~C.~Ni\c{t}u]{
Michael J. Keith$^{1}$\thanks{E-mail: mkeith@pulsarastronomy.net} and
Iuliana C. Ni\c{t}u$^{1}$
\\
$^{1}$Jodrell Bank Centre for Astrophysics, Department of Physics \& Astronomy, The University of Manchester, M13 9PL, Manchester, UK\\
}

\date{Accepted XXX. Received YYY; in original form ZZZ}

\pubyear{2023}

\begin{document}
\label{firstpage}
\pagerange{\pageref{firstpage}--\pageref{lastpage}}
\maketitle

\begin{abstract}
Timing noise in pulsars is often modelled with a Fourier-basis Gaussian process that follows a power law with periodic boundary conditions on the observation time, $T_\mathrm{span}$.
However the actual noise processes can extend well below $1/T_\mathrm{span}$, and many pulsars are known to exhibit quasi-periodic timing noise.
In this paper we investigate several adaptions that try to account for these differences between the observed behaviour and the simple power-law model.
Firstly, we propose to include an additional term that models the quasi-periodic spin-down variations known to be present in many pulsars.
Secondly, we show that a Fourier basis of $1/2T_\mathrm{span}$ can be more suited for estimating long term timing parameters such as the spin frequency second derivative (F2), and is required when the exponent of the power spectrum is greater than $\sim 4$. 
We also implement a Bayesian version of the generalised least squares `Cholesky' method which has different limitations at low frequency, but find that there is little advantage over Fourier-basis methods.
We apply our quasi-periodic spin down model to a sample of pulsars with known spin-down variations and show that this improves parameter estimation of F2 and proper motion for the most pathological cases, but in general the results are consistent with a power-law model.
The models are all made available through the \textsc{run\_enterprise} software package.

\end{abstract}




\section{Introduction}
Pulsar timing relies on the comparison of the observed time of arrival (ToA) of a pulse from a pulsar with a parametric model of the pulsar's spin, astrometric and other parameters.
The difference between the observed ToA and the model for a given observation is termed the residual, and in an ideal case the post-fit residual would be white noise well described by the uncertainty on the ToA.
However, it is well known that pulsar residuals often exhibit excess noise with both a white noise component \citep{oslowski11,Parthasarathy21}, independent from observation to observation, and red noise processes which are correlated on timescales much longer than the typical observing cadence \citep{cordes85,hobbs2010,asj+19}.
For many pulsars the red noise processes are very significant and can deviate from the model by many rotations.

Estimating the power-spectral density of the residuals reveals that the red noise typically appears well modelled by a power-law process in the Fourier domain \citep{reardon16,asj+19}.
The exponent of these power laws can be very steep, which hampers simple techniques for estimation of the power-spectral density, and typically requires using a method of pre-whitening the data using either a differential method or using the covariance matrix.
\citet{coles} showed that common fitting methods, such as weighted least-squares can be biased by this red noise, even when using simple methods to deal with the red noise by fitting polynomials or sinusoids.
The same paper also demonstrates a method to estimate the covariance matrix of the residuals from the estimate of the power-spectral density, which can then be used with generalised least-squares (GLS) to solve for the other parameters in the presence of red noise.
Although this so-called `Cholesky' method is effective, the iterative nature of requiring the covariance function of the data to properly estimate the power-spectral density used to model the covariance function means that it is possible to find a false peak in the overall likelihood.
Bayesian techniques were proposed to solve for the data covariance function and the model parameters simultaneously, analytically integrating over less critical parameters in the timing model \citep{vl13,temponest}.

The long-term fluctuations of the generally more stable millisecond pulsars are also expected to contain red noise terms caused by a stochastic background of gravitational waves.
Detection of these waves in pulsar timing data is a key objective of several large international collaborations \citep{epta_gw6,ppta_dr2,ng_12GW,inpta,meertime_pta}.
This has necessitated the optimisation of the Bayesian algorithms to reduce the very large computational complexity of solving a model containing perhaps 100 pulsars simultaneously.
This led to the development of modelling the red noise as a Fourier-basis Gaussian process which can greatly reduce the number of computations required to fit the red noise models \citep{temponest_physrevd}.
These methods have been widely applied to pulsar data for gravitational wave detection in codes such as \textsc{temponest} \citep{temponest} and \textsc{enterprise} \citep{enterprise}.

These models are also recently being more widely applied to the timing of the larger population of pulsars (e.g. \citealp{asj+19}).
However, these algorithms are optimised for millisecond pulsars where the scale of the red noise means that only a small number of Fourier components are required to fully model the noise.
Furthermore, one might question if the periodic nature of a Fourier-basis Gaussian process may impact the measurement of long-term spin parameters such as the spin frequency second derivative that is an important measurable when considering the long-term evolution of pulsars.
More importantly, although the red noise in pulsars is often well fit by a power-law, we have observed quasi-periodic spin evolution in a growing number of pulsars (e.g. \citealp{hobbs2010,asj+19}).

There have been some attempts to include these quasi-periodic variations in the timing noise model.
For example studies of a sample of pulsars regularly timed with the Parkes radio telescope have been searched for periodic modulations in the timing residuals \citep{khjs16,asj+19}.
In these cases, the method employed was to add a sinusoidal component to the timing model with a free amplitude and periodicity, and use likelihood or Bayesian evidence ratios to test the favourability of a periodic model.
Similarly, recent searches for planetary companions around pulsars on the Jodrell Bank pulsar timing archive demonstrated that periodic timing variations are detectable in a large number of pulsars \citep{nks+22}.
However, although these methods are effective at detecting some quasi-periodic variability, especially in cases where the periodicity is relatively constant, they do not fully model the quasi-periodic variations and therefore may not be optimal for parameter estimation.
In this work we build on these ideas by developing a quasi-periodic term for the power-law timing model.

Young pulsars also often show a significant measurement $\ddot{\nu}$, the long-term frequency second derivative (F2).
Estimation of F2 is important for the understanding of the long-term evolution of pulsars, and is important for understanding the pulsar braking index and in the recovery from pulsar glitches \citep{2011ApJ...741L..13E,2015MNRAS.446..857L,2021MNRAS.508.3251L}, as well as having implications for high precision pulsar timing experiments \citep{xkbs19}.
The F2 parameter appears as a cubic term in the residuals, and hence is most sensitive on the longest timescales, which are also those where the timing noise dominates.
Therefore the choice of timing noise model may be important for estimating the magnitude of F2, and particularly for quantifying the significance of any measurement.
Particularly, the important question is if the observed F2 is consistent with being the low-frequency extension of the timing noise observed on shorter timescales, or if it reflects an additional process such as glitch recovery or the long term braking of the pulsar.

In this paper we will attempt to address two questions regarding the application of the current Fourier-basis Gaussian process models on the canonical (i.e. non-millisecond) pulsar populations.
\begin{itemize}
    \item Does the periodic boundary condition of Fourier basis models affect the ability to measure F2 for young pulsars? If so can this be mitigated by changing the lowest basis frequency?
    \item Is there any advantage to modelling the pulsar timing noise with a model containing a quasi-periodic component, rather than the pure power-law model traditionally used? 

\end{itemize}
We also take the opportunity to implement a Bayesian version of the \citet{coles} model within the \textsc{enterprise} framework and confirm the results of \citet{temponest} that the Fourier-domain Gaussian process model performs equally as well as the GLS method for the estimation of pulsar timing parameters.

\section{Timing noise modelling}
\label{sec:tn_general}
We make use of the \textsc{enterprise} framework for the Fourier-basis Gaussian process models.
A complete description of this method can be found in \citet{temponest_physrevd} and \citet{temponest}.
In brief, we define a model of the Fourier-domain power-spectral density (PSD) and use this to constrain the amplitude of a harmonic series of sinusoids, i.e. a Fourier basis.
The PSD is typically modelled as a power-law, parameterised by the log-amplitude $\mathrm{log}_{10}(A)$ and the spectral exponent $\gamma$.
For a stochastic Gaussian process, the underlying PSD is equal to the variance in the amplitude of the corresponding sinusoid, and so we can fit for harmonically related sinusoids with a Gaussian prior with variance defined by the model PSD.
In practice this constrained fit is analytically marginalised over, making the method very computationally efficient.

The GLS approach assumes that the noise is drawn from a Normal distribution with zero mean and covariance given by a covariance matrix $\mathbf{C}$.
We typically estimate $\mathbf{C}$ by the Wiener-Khinchin theorem, which states that covariance as a function of lag is the Fourier transform of the PSD.
\citet{coles} propose fitting a power-law model to estimates of the PSD computed using a periodogram analysis of the pulsar data.
However, this estimation of the PSD may depend on the choice of pulsar timing parameters, which in turn may depend on the choice of $\mathbf{C}$.
This can be somewhat addressed by an iterative approach, but this still does not allow for the uncertainty in the PSD model to be factored into the uncertainty on other fit parameters.
The Bayesian approach is to fit the PSD hyperparameters (the amplitude, $A$ and exponent, $\gamma$, of the power-law) at the same time as solving for other parameters of interest, whilst using the $\mathbf{C}(A,\gamma)$ to evaluate the likelihood, and GLS to analytically marginalise over any other linear parameters.

It is worth noting that for both cases, we cannot strictly model a pure power-law.
For the GLS method, the integral to compute $\mathbf{C}$ is not finite for a pure power-law, and so a corner frequency, $f_\mathrm{c}$ is introduced, below which the PSD flattens, typically chosen to be of order $1/T_\mathrm{span}$.
The Fourier basis model must also make a choice of the set of harmonically related frequencies to use.
Typically for a dataset of length $T_\mathrm{span}$ the lowest frequency used is of order $f_\mathrm{low} = 1/T_\mathrm{span}$.
The number of harmonics, $N_\mathrm{harm}$ used is also finite, and the natural choice is to ensure that the PSD at the frequency $f_\mathrm{high}=N_\mathrm{harm} f_\mathrm{low}$ is dominated by white noise.
Unlike the GLS approach, the Fourier basis model does not flatten below $f_\mathrm{low}$, but rather the use of a Fourier basis imposes periodic boundary conditions.
This means that the Fourier basis red noise model must be periodic over a window of $1/f_\mathrm{low}$.
Clearly the intrinsic timing noise has no knowledge of our observing span, and therefore any timing noise longer than this timescale will leak into the fit parameters sensitive to the longest timescales.
Usually it is assumed that this only has a small perturbation on the estimation of the long-term spin, F0, and spin derivative, F1, parameters.
In effect it means that F0 and F1 represent the average spin frequency and spin-down frequency over the observing timespan rather than estimates of the long-term spin of the pulsar, but this usually has little impact on their interpretation.

\subsection{Modelling quasi-periodic variations in spin down}
\label{sec:qp_model}
The widely used timing noise models typically use a power-law model for the PSD model, with some use of multiple or broken power-laws.
However there is good evidence that many pulsars show quasi-periodic behaviour, which may not be well modelled by a power-law.
Here we propose a modification to the existing PSD model to include a quasi-periodic term.
We choose to do this in the Fourier domain to re-use the existing frameworks, but equivalent time-domain covariance kernels could also be constructed.
We have chosen to model the quasi-periodic process with a Gaussian function centred on the fundamental frequency and  harmonics of a periodic process.
Motivated by the observed power spectra of pulsars exhibiting quasi-periodic timing noise (particularly PSRs B1828$-$11 and B1540-06), the Gaussian functions broaden and decrease in amplitude at higher harmonics.
Particularly the integral of each Gaussian decays exponentially, and the Gaussians maintain a constant fractional bandwidth.
This choice is largely arbitrary, but is similar in intent to the relatively common choice of a cosine or sine-squared kernel multiplied by a Gaussian envelope in the time domain.

The quasi-periodic process described in \citet{lyne2010} appears in the spin-frequency derivative, $\dot{\nu}$, of the pulsar.
Therefore we begin by considering a quasi-periodic Fourier-domain Gaussian process in $\dot{\nu}(t)$, with a PSD composed of a train of Gaussian components with integral of the $k$th harmonic decaying like $\exp{(-(k-1)/\lambda)}$, where $\lambda$ is a free parameter that effectively describes the number of significant harmonics in the data. 
Specifically, we define the spectral shape of our model by
\begin{equation}
\label{qp_function}
    q(f,f_\mathrm{qp},\sigma,\lambda) = \sum_{k=1}^N \frac{\exp{(-(k-1)/\lambda)}}{k} \exp{ \left(\frac{-(f-kf_\mathrm{qp})^2}{2k\sigma^2} \right) },
\end{equation}
where $\sigma$ is the width of the Gaussian function and $f_\mathrm{qp}$ is the fundamental frequency ($k=1$ harmonic) of the quasi-periodic process. 
Our analysis requires a model of the PSD in the residuals, $r(t)$, which is related to $\dot{\nu}$ by $\dot{\nu}(t) \propto \ddot{r}(t)$.

Hence, to convert our model of $\dot{\nu}(t)$ to the effect on the residuals we integrate twice with respect to time.
In the Fourier domain we only need to integrate the Fourier basis functions with respect to time, which has the effect of multiplying the power spectrum by $f^{-4}$, ignoring constant scaling factors.
The resulting power spectrum is therefore proportional to $f^{-4} q(f,f_\mathrm{qp},\sigma,\lambda)$.

The choice of a Gaussian function for the spectral shape of our quasi-periodic process means that the power decays fairly quickly away from the peak of the quasi-periodic function.
In practice we find that this is not sufficient to model the observed spin variations and the observed PSD appears more like a quasi-periodic process on top of a power-law model.
Some of this may be because the $\dot\nu$ variation process has intrinsically broader `tails' than a Gaussian function.
This might be observed if the $\dot\nu$ variation process is behaving more like a random switching of $\dot\nu$ and hence a random walk in $\nu$. 
This would lead to a $f^{-2}$ PSD in $\nu$ and hence a $f^{-4}$ power-law process in $r(t)$.
One approach for modelling this would be to replace the Gaussian function with a function with wider tails (e.g. the probability density function of the Cauchy distribution), but we find in practice it is simpler to add an additional power-law term to the model.
This also has the advantage that it allows us to separate the `purely' quasi-periodic component from the excess power-law noise and capture both the excess $\dot\nu$ noise as a process with spectral exponent of $-4$, but also allows us to model a wider range of noise processes with different spectral indices.
Indeed, we will see that in practice $\gamma$ is significantly divergent from $-4$ for some pulsars.
The practical implementation of this model is described in Section \ref{sec:nm}.

\subsection{Implementation of the noise models}
\label{sec:nm}
\subsubsection{Fourier basis Gaussian Process}
For the Fourier-basis Gaussian process models we use the model of \citet{temponest} as implemented in \textsc{enterprise}.
The power-law noise model is defined by a one-sided power spectral density
\begin{equation}
    P_\mathrm{pl}(f) = \frac{A_\mathrm{pl}^2}{12\pi^2} \left(\frac{f}{f_\mathrm{yr}}\right)^{-\gamma} f_\mathrm{yr}^{-3},
    \label{ppl_ent}
\end{equation}
where $f_\mathrm{yr}$ is a frequency of 1 per year. The hyperparameters are the spectral exponent $\gamma$, and log-amplitude $\log_{10}(A_\mathrm{pl})$.

There are numerous ways to model quasi-periodic variations, and in this work we have chosen the model described in Section \ref{sec:qp_model}. Specifically our model for quasi-period variations in $\dot\nu$ has a power-spectral density given by
\begin{equation}
    P_\mathrm{qp}(f) = R_\mathrm{qp}P_\mathrm{pl}(f_\mathrm{qp}) q(f,f_\mathrm{qp},\sigma,\lambda)\left(\frac{f}{f_\mathrm{yr}}\right)^{-4},
    \label{ppl_qp}
\end{equation}
where $R_\mathrm{qp}$ is the ratio of quasi-periodic noise to red noise at the central frequency $f_\mathrm{qp}$, and $q(f,f_\mathrm{qp},\sigma,\lambda)$ is the function describing the spectral shape of our quasi-periodic process, defined in Equation~\ref{qp_function}.

This model consists of $N$ harmonically related Gaussian functions of width $\sigma$, which decay with an exponential scale $\lambda$. The hyperparameters of the quasi-periodic model are $\log_{10}{(R_\mathrm{qp})}$, $f_\mathrm{qp}$, $\sigma$ and $\lambda$.
 The final red noise model including the quasi-periodic $\dot\nu$ term is $P_\mathrm{qp}(f)+P_\mathrm{pl}(f)$, and so contains the 6 hyperparameters from both red noise models.

In order to prevent $P_\mathrm{qp}$ growing at very low frequencies, where the Gaussian function decays slower than $f^{-4}$, a low-frequency cut-off is applied such that $P_\mathrm{qp} = 0$ for $f<f_\mathrm{cut}$.
We choose $f_\mathrm{cut}$ to be the local minima value of $P_\mathrm{qp}$ for $N=1$ (i.e. for only the lowest frequency harmonic) and hence can be determined by differentiation to be
\[
f_\mathrm{cut} = \frac{1}{2} \left(f_\mathrm{qp} - \sqrt{(f_\mathrm{qp}^2 - 16\sigma^2)}\right).
\]

\subsubsection{Generalised Least Squares}
When directly computing the covariance function for GLS fitting we cannot use a pure power-law as this diverges as the frequency approaches zero, so we adapt the model proposed by \citet{coles}, where the two-sided power-spectral density is given by,
\begin{equation}
\label{ppl_coles}
    P_\mathrm{pl}(f) = P_\mathrm{ref} \left(1+\left[\frac{f}{f_\mathrm{c}}\right]^2\right)^{-\gamma/2}\left(\frac{f_\mathrm{c}}{f_\mathrm{ref}}\right)^{-\gamma},
\end{equation}
where $f_\mathrm{c}$ is the corner frequency below which the spectrum turns over and $f_\mathrm{ref}$ is the reference frequency at which $P_\mathrm{ref}$ is defined.
This modification allows us to sample the power-spectral density at a meaningful scale, rather than at $f_\mathrm{c}$ which may correspond to a timescale well beyond the observing span.
This simplifies to the  Coles et al. model for $f_\mathrm{c}=f_\mathrm{ref}$, and approaches Equation \ref{ppl_ent} for $f_\mathrm{ref}=f_\mathrm{yr}$ and small values of $f_\mathrm{c}$.
Specifically, for $f_\mathrm{c}$ significantly smaller than $1/T_\mathrm{span}$, conversion between Fourier basis Lentati et al. models (as output by \textsc{enterprise} or \textsc{temponest}) and Coles et al. GLS models (as used in \textsc{tempo2}) can be made by making use of the conversion
\begin{equation}
    2P_\mathrm{ref} =  \frac{A_\mathrm{pl}^2}{12\pi^2} \left(\frac{f_\mathrm{ref}}{f_\mathrm{yr}}\right)^{-\gamma} f_\mathrm{yr}^{-3},
\end{equation}
where the factor of $2$ converts from one-sided to two-sided power-spectral density.

It is of course possible to also implement the quasi-periodic model within the GLS framework.
The quasi-periodic model remains unchanged, except being scaled by $P_\mathrm{pl}(f)$ from Equation \ref{ppl_coles} rather than Equation \ref{ppl_ent}.

\subsubsection{White noise}
In addition to red noise, pulsars are also seen to exhibit excess white noise \citep{oslowski11,Parthasarathy21}.
Hence for the analysis in this paper we expand each of our noise models to also include the widely used `EFAC' and `EQUAD' white noise parameters that linearly scale and add in quadrature to the formal ToA uncertainty respectively.
We follow the convention in \textsc{temponest} \citep{temponest} such that the output ToA error relates to the input ToA error, $\sigma_\mathrm{toa}$, by
\begin{equation}
    \sigma_\mathrm{out} = \sqrt{(\mathrm{EFAC}\times\sigma_{\mathrm{toa}})^2 + \mathrm{EQUAD}^2}.
\end{equation}

\subsubsection{The four models}
We therefore have four models, two using the Fourier basis and two using the GLS, which are summarised in Table \ref{models_table}. These have been implemented as part of a single pulsar Bayesian toolkit \textsc{run\_enterprise} \citep{run_enterprise}, which utilises the \textsc{enterprise} framework and the pulsar timing package \textsc{tempo2} as exposed by the python interface \textsc{libstempo}.
\textsc{run\_enterprise} has been developed to allow fitting and comparison of a wide combination of pulsar noise models and timing parameters using a range of Bayesian samplers.
In this work we sample the hyperparameters using \textsc{multinest} \citep{multinest1}, via the python interface \textsc{pymultinest} \citep{pymultinest}.

\begin{table}
    \centering
       \caption{The four models used in this work}
    \label{models_table}
    \begin{tabular}{cccc}
    Model & Method & Type & Equations\\
    \hline
        PF & Fourier Basis & Power-law & (\ref{ppl_ent}) \\
        PG & GLS & Power-law & (\ref{ppl_coles})\\
        QF & Fourier Basis  & Power-law + Quasi-periodic& (\ref{ppl_ent}) + (\ref{ppl_qp})\\
        QG & GLS  & Power-law + Quasi-periodic& (\ref{ppl_coles}) + (\ref{ppl_qp})\\
    \end{tabular}
 
\end{table}

\section{Low frequency cut-off and periodic boundary conditions}
\label{sec:lowfreq}
\begin{figure}
    \centering
    \includegraphics[width=\columnwidth]{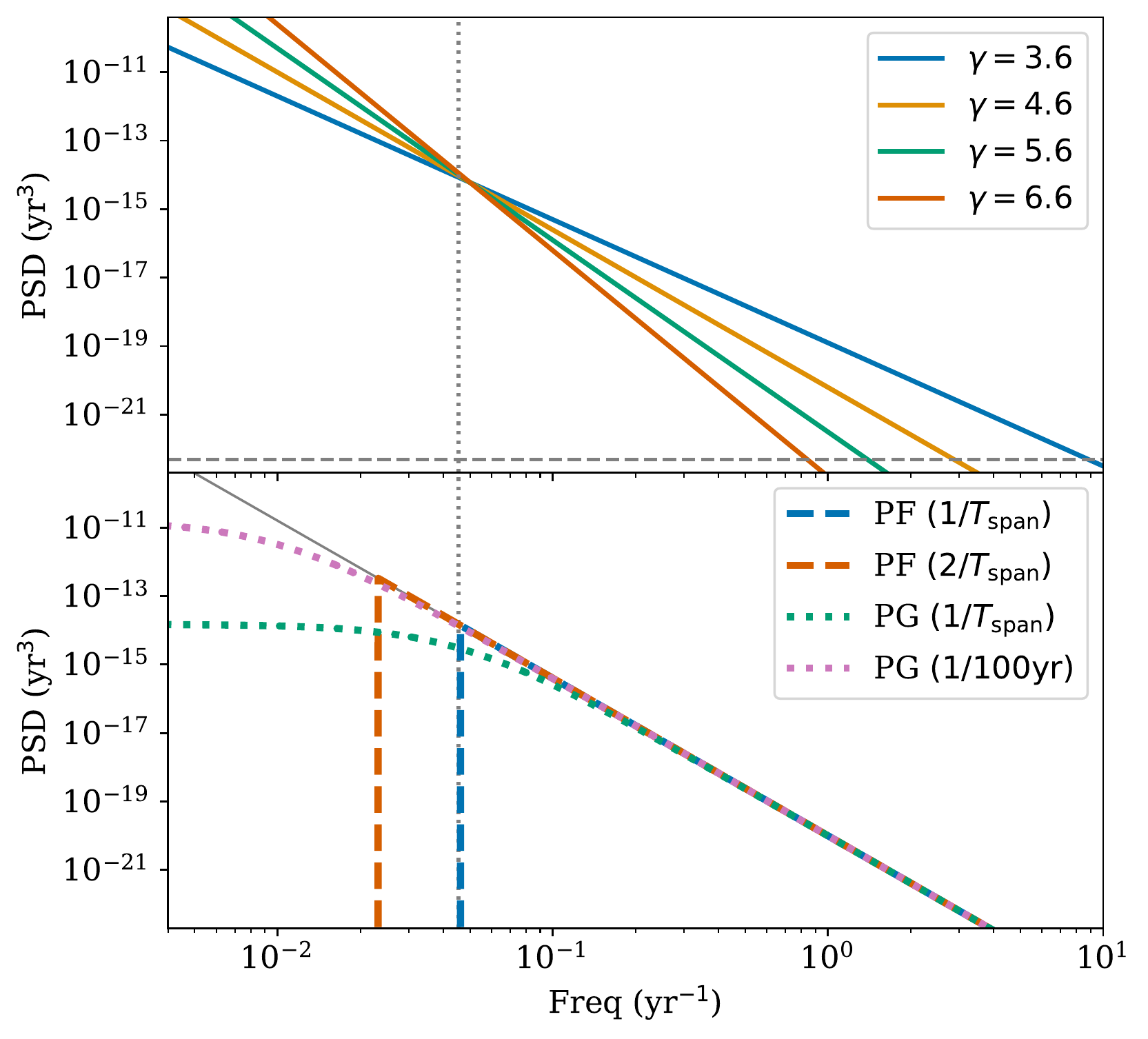}
    \caption{\label{fig:pl}Upper panel: The injected power-law red noise for $\gamma=3.6$, $4.6$, $5.6$ and $6.6$. The dotted vertical line is at $1/T_\mathrm{span}$ and the dashed gray line shows the white noise PSD. Lower Panel: The behaviour of the four different `power-law' models used for $\gamma=4.6$. Dashed curves are for the PF model with $f_\mathrm{min}=1/T_\mathrm{span}$ and $f_\mathrm{min}=1/2T_\mathrm{span}$ and dotted curves are for the PG model with $f_\mathrm{c} = 1/T_\mathrm{span}$ and $f_\mathrm{c} = 1/100 \mathrm{yr}^{-1}$. The dotted vertical line is at $1/T_\mathrm{span}$ and the solid grey line is a pure power-law.}
    
\end{figure}

As we note in Section \ref{sec:tn_general}, even if the intrinsic spin-noise is a power-law over the observation time-span, we know that the power-law cannot extend to zero frequency as the PSD must remain finite whilst the power-law diverges. The two types of models we use here behave differently at the lowest frequencies. The Fourier basis models define a lowest frequency that is both a cut-off for the power-law and also defines the time window for the periodic boundary conditions imposed by the Fourier basis on the red noise model.
The GLS models define a corner frequency at which the model turns over and hence can have a finite integral.
It is generally assumed that the effects of the frequencies below $1/T_\mathrm{span}$ are absorbed in small changes in the F0 and F1 parameters, but this will also affect the estimation of F2 to some extent. Although these models have been tested extensively during their development, the impact on the F2 parameter was not the focus of these tests, yet F2 is an important parameter for studying the long-term spin evolution of pulsars.
To investigate the impact of the choice of model on F2 estimation we perform a simple simulation.
We generate 224 simulated ToAs over a timespan of 22 yr, with cadence and measurement errors and pulsar parameters based on the actual Lovell telescope observations of PSR J1909+0912.
In addition to white noise, we inject a power-law spin noise that extends to a frequency of $1/200$ yr$^{-1}$, i.e. the longest periodicity is $\sim 9$ times $T_\mathrm{span}$. We run four sets of simulations, with red-noise spectral exponents of 3.6, 4.6, 5.6 and 6.6 to reflect the typical range of values we observe in the young pulsars where F2 is likely to be measured \citep{asj+19}. The amplitude of the red noise is chosen such that it produces a root-mean-square (rms)  residual similar to the real pulsar, which is dominated by the red noise, and scaled between the four exponents such that the power at $1/T_\mathrm{span}$ is approximately the same. The four injected power-laws are shown in the upper panel of Figure \ref{fig:pl}.
We compare four approaches to estimating the red noise and F2.
\begin{itemize}
    \item PF model with the lowest frequency at $1/T_\mathrm{span}$, as typically used for gravitational wave studies;
    \item PF model with lowest frequency at $1/2T_\mathrm{span}$ to reduce requirements for periodic boundary conditions;
    \item PG model with $f_\mathrm{c} = 1/T_\mathrm{span}$, as suggested by \citet{coles};
    \item PG model $f_\mathrm{c} = 1/100 \mathrm{yr}^{-1}$, a commonly used alternative for when it is felt that $f_\mathrm{c} \ll 1/T_\mathrm{span}$.
\end{itemize}
There are of course many other choices that could be made, but we feel that these four explore a reasonable range of commonly used models.
An example of these four models is shown in the lower panel of Figure \ref{fig:pl}.
It is worth noting that the PG models diverge from a power-law well before reaching $f_\mathrm{c}$, and hence the power at $1/T_\mathrm{span}$ is very different for the $f_\mathrm{c} = 1/T_\mathrm{span}$ model compared to the other three models, which only significantly diverge from the power-law at frequencies below  $1/T_\mathrm{span}$.

\subsection{Results from simulations}
\label{pl_results}
For each choice of exponent, we run 40 realisations of the simulation, and fit for the noise hyperparameters and F2, marginalising over F0, F1 and the position of the pulsar.
The initial parameters for the fitting were slightly perturbed from the simulated values.
For a parameter $p$, in this case F2, we have a truth value $p_\mathrm{truth}$, the best-fit value, $\hat{p}_i$, and error, $e_{p,i}$, of the $i$th realisation of the simulation.
We compute the average deviation from the injected value, 
\begin{equation}
\Delta_{p} = \frac{1}{N}\sum\limits_{i=0}^{N} \left(\hat{p}_i-p_\mathrm{truth}\right),
\end{equation}
and $\sigma_p^2$ which is the variance of the $\hat{p}_i$ values.
Note the distinction between $e_{p,i}$, the formal error returned by the fit algorithm, and $\sigma_p$ which is derived from the observed variation between realisations of the simulation.
Using these, we can derive metrics for the quality of the results obtained.
Firstly we take $\sigma_p/\sigma_{p,\mathrm{min}}$, rms in the results normalised to the model with the lowest rms.
This is the factor by which the actual scatter in the parameter estimates increases over the `best' model.
Next we consider how well the fit errors model the scatter.
Since the values of $e_{p,i}$ are consistent from realisation to realisation, we simply consider the mean error estimate $\overline e_p$.
If the errors are well estimated, $\overline e_p /\sigma_p$, will be unity.
A greater value indicates errors are typically over-estimated, and a smaller value indicates errors are typically under-estimated.
Finally we consider an estimate of the normalised bias, $\Delta_{p} / \sigma_p$, which should be consistent with zero.
The results for fitting F2 in these models are given in Table \ref{simparamtab2}.

For $\gamma=3.6$, the PF models and the PG model with $f_\mathrm{c} = 1/100\mathrm{yr}$ all return reasonable values. The PG model with $f_\mathrm{c} = 1/T_\mathrm{span}$ tends to underestimate the uncertainties, as might be expected since in this model the PSD starts to turn over well before the frequency range influenced by the F2 fit.
As the spectral exponent increases, the PF model with lowest frequency at $1/T_\mathrm{span}$ and the PG model with $f_\mathrm{c}=1/T_\mathrm{span}$ begin to significantly underestimate the uncertainties. The PG model with $f_\mathrm{c} = 1/100\mathrm{yr}$ also starts to underestimate the uncertainty as the red noise becomes steeper. The PF model with lowest frequency at $1/2T_\mathrm{span}$ fairs better, consistently returning reliable uncertainties at all tested values of $\gamma$.

However, it is worth noting that we do not know what happens to the red-noise PSD at frequencies below our observing span, and with our longest datasets at $\sim 50$ years, we have no measurements for frequencies close to $1/100\mathrm{yr}$.
We should therefore always keep in mind that the measurements of F2 (and F1 and F0) depend on our choice of what happens to the red noise at frequencies below $1/T_\mathrm{span}$, and hence we may under (or over) estimate the significance of F2 even with a very small choice of $f_\mathrm{c}$ or $f_\mathrm{low}$.
Nevertheless we believe these simulations support our assertion that using the PF model with lowest frequency at $1/2T_\mathrm{span}$ significantly improves the uncertainty estimate on F2 under the influence of steep red noise.
Using the PG model avoids any problems with the periodic boundary conditions, but the uncertainty in F2 depends strongly on the choice of $f_\mathrm{c}$.
Given the additional computational cost of the PG model, especially for small values of $f_\mathrm{c}$, we therefore would recommend using the PF model (or QF model if appropriate) with lowest frequency at $1/2T_\mathrm{span}$ when studying the long-term evolution of pulsars.

\begin{table}
    \centering
        \caption{Statistics of parameter estimates from 40 realisations of simulated observations with power-law red noise. Parameters in the columns are defined in Section \ref{pl_results} and from left to right represent the scatter, error over-estimation factor, and bias in the fitting.}
    \label{simparamtab2}
    \begin{tabular}{clcccc}
    $\gamma$  & Model & $\sigma_p/\sigma_{p,\mathrm{min}}$ & $\overline e_p /\sigma_p$ &  ${\Delta_{p}} / \sigma_p$ \\
        \hline

3.6   & PF ($1/T_\mathrm{span}$) &     1.0(1) &    1.4(2) & 0.0(2)\\
3.6   & PF ($1/2T_\mathrm{span}$)  &     1.0(1) &    1.2(1) & 0.0(2)\\

3.6   & PG  ($1/T_\mathrm{span}$)  &     1.0(1) &   0.7(1) & 0.0(2)\\
3.6   & PG  (1/100yr)    &     1.0(1) &    1.1(1) & 0.0(2)\\
\hline
4.6   & PF ($1/T_\mathrm{span}$) &     1.1(1) &    1.0(1) & 0.1(2)\\
4.6   & PF ($1/2T_\mathrm{span}$)  &     1.0(1) &    1.2(2) & 0.1(2)\\

4.6   & PG  ($1/T_\mathrm{span}$)  &     1.1(1) &   0.6(1) & 0.1(2)\\
4.6   & PG  (1/100yr)    &     1.0(1) &    1.0(1) & 0.1(2)\\
\hline

5.6   & PF ($1/T_\mathrm{span}$) &     1.0(1) &   0.37(4) & -0.3(2)\\
5.6   & PF ($1/2T_\mathrm{span}$)  &     1.0(1) &    1.1(1) & -0.3(2)\\
5.6   & PG  ($1/T_\mathrm{span}$)  &     1.1(1) &   0.33(4) & -0.3(2)\\
5.6   & PG  (1/100yr)    &     1.0(1) &   0.61(7) & -0.3(2)\\
        \hline
6.6   & PF ($1/T_\mathrm{span}$) &     1.1(1) &   0.09(1) & -0.1(2)\\
6.6   & PF ($1/2T_\mathrm{span}$)  &     1.0(1) &   0.74(9) & -0.1(2)\\
6.6   & PG  ($1/T_\mathrm{span}$)  &     1.2(1) &   0.16(2) & -0.1(2)\\
6.6   & PG  (1/100yr)    &     1.1(1) &   0.27(3) & -0.1(2)\\
    \end{tabular}

\end{table}

\section{Quasi-periodic spin-down}

\subsection{Simulations}
\label{sec:simqp}
In order to demonstrate the limitations of using power-law models for pulsars which exhibit quasi-periodic spin-down variations, we construct another simple simulation.
The simulated pulsar switches between two spin-down states corresponding to a change of $\dot\nu$ by $1\%$, with each `high' state lasting for $162\pm 10$ days and each `low' state lasting for $440\pm10$ days.
We again generate 224 ToAs over $22\,\mathrm{yr}$, and based on actual Lovell telescope observations of PSR J1909+0912.
This represents a highly, but not perfectly, periodic variation in the pulsar spin, similar to a pulsar such as PSR B1828$-$11 or B1540$-$06.
In order to test the effect of the mis-match between model and data we consider two outputs - firstly the estimates of the power-spectral density of the data, and secondly the estimates of the pulsar parameters.

\begin{figure}
    \centering
    \includegraphics[width=\columnwidth]{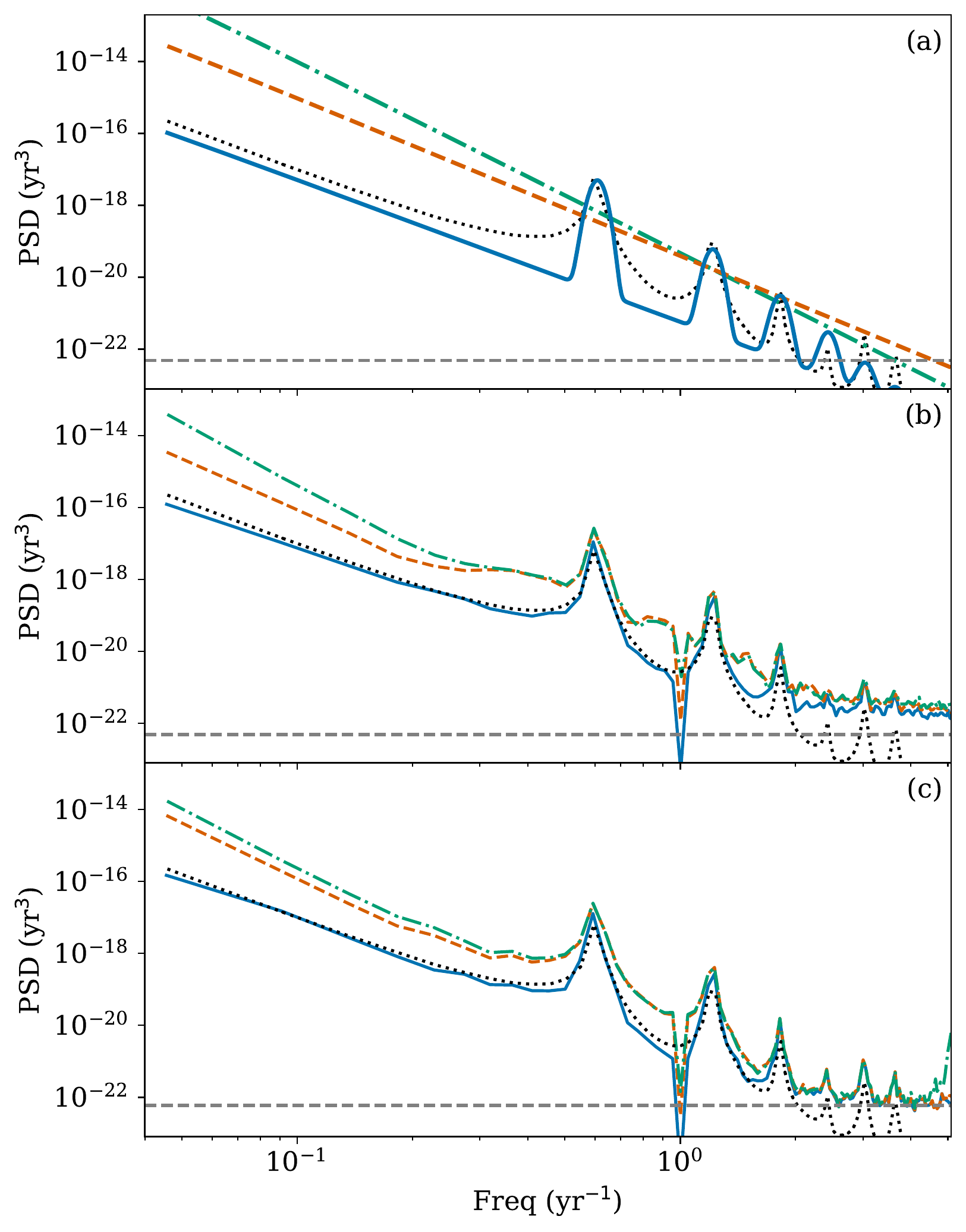}
    \caption{\label{simulated_mean_psd} Power spectra derived from the simulations in Section \ref{sec:simqp}. (a) The maximum-likelihood model power-spectrum for a single realisation of ToAs from the simulation of quasi-periodic $\dot{\nu}$ variations on real pulsar sampling. (b) The mean value of the PSD estimated from 20 realisations of the simulation with irregular sampling. (c) The mean value of the PSD estimated from 20 realisations of the same simulation but with regular sampling. In each panel the solid lines are for the QG, dotted lines are for the PG model and dot-dashed lines are for the PF model. The QF model is not shown as it almost exactly overlaps with the QG model. The dotted spectrum is the average of the 1024 realisations of the simulation computed directly using the DFT on the generated $\dot\nu$ sequence. The dashed horizontal line is the mean PSD of the injected white noise only.}
\end{figure}

We use the \textsc{cholspectra} plugin of \textsc{tempo2} to estimate the one-sided power-spectral density based on the maximum likelihood models from each realisation of the simulation.
The stochastic $\dot\nu$ switching model does not lend itself to an analytic model of the PSD, but we can estimate the expected PSD under ideal conditions by simulating a large number of uniformly sampled realisations of the injected signal.
To avoid the spectral leakage prevalent when taking a periodogram of processes that fall faster than $f^{-2}$, we take the discrete Fourier transform of $2^{14}$ uniformly spaced samples of each realisation of the injected $\dot\nu$ time-series, and multiply the resulting spectrum by $f^{-4}$.
This should be equivalent to integrating twice in the time domain and hence recover the model spectrum for the residuals, though we caution that aliasing may lead to excess high-frequency noise that is then amplified if the spectrum does not decay rapidly enough at high frequencies.

Figure \ref{simulated_mean_psd} shows the average of $2^{10}$ realisations of the PSD estimated in this way, overlaid on the maximum likeihood model and an average of the PSD from \textsc{cholspectra}.
As might be expected, for such strongly periodic signals, the pure power-law models primarily capture the peaks in the spectra, and hence significantly overestimate the PSD at frequencies away from the harmonics of the quasi-periodic signal.
In panel (b) of Figure \ref{simulated_mean_psd} we show the average PSD from the 20 realisations of the simulation.
Here we see that on average the use of power-law models leads to a substantial over-estimation of the PSD, especially at low frequencies where it can be out by more than 2 orders of magnitude.
All models also overestimate the PSD at high frequencies, flattening well before reaching the injected white noise level, giving the appearance of excess white noise.
We attribute this high frequency excess to spectral leakage due to a combination of the irregular sampling of the data and the fact that the model does not perfectly capture the underlying PSD.
This effectively limits the dynamic range available in the power-spectrum with irregular sampling.
To test this hypothesis, we repeated the simulation with the same number of ToAs uniformly sampled across the same timespan, with the same white noise and quasi-periodic $\dot\nu$ switching.
The average PSD from the 20 simulations with uniform sampling is shown in Figure \ref{simulated_mean_psd}(c), and we can see that the estimated PSD correctly approaches the white noise level, but the low-frequency behaviour is largely unchanged.

This apparent excess white noise in irregular sampling may also be seen in real pulsar observations due to similar mismatches between the model and underlying noise process, though it is not clear how to distinguish this from excess white noise intrinsic to the pulsar.
We note that in the case of irregular sampling, the spectral estimates at the high frequencies are highly correlated - indeed this must be the case since the effective Nyquist frequency is much higher than for the regularly sampled case yet the number of input data points has not increased.

In this simulation the PSD close to the periodicity is approximately $10^3$ times the power-law component, however it is important to note that this is only discernible when the PSD is formed using the quasi-periodic model.
We also note that the simulation does not contain a separate power-law component, but rather the quasi-periodic $\dot\nu$ variations intrinsically cause a power-law red noise, at least below the fundamental frequency of the quasi-periodic process.

For parameter estimation, we focus on three parameters:
the long-term frequency second derivative, F2, and the two proper-motion parameters PMRA and PMDEC.
These parameters have interest to the pulsar astronomer, and have been shown they can be biased if incorrectly dealing with red noise \citep{coles}. The results of fitting these parameters in 20 simulations are summarised in Table \ref{simparamtab}, using the same metrics as in Section \ref{pl_results}.
In all cases the quasi-periodic models are both a factor of 3--5 more precise, and have better error estimates.
In practice the real pulsars have a wide range of observed periodicities, and the benefit of using a quasi-periodic model will be highly situational, depending on both the periodicity, amplitude and purity of the observed quasi-periodic process.

\begin{table}
    \centering
        \caption{Statistics of parameter estimates from 20 realisations of simulated observations with $\dot\nu$-switching. Columns are as in Table \ref{simparamtab2} and defined in Section \ref{pl_results}.}
    \label{simparamtab}
    \begin{tabular}{cccccc}
    Parameter & Model & $\sigma_p/\sigma_{p,\mathrm{min}}$ & $\overline e_p /\sigma_p$ &  ${\Delta_{p}} / \sigma_p$ \\

    \hline
F2    & PF               &     5.0(5) &     13(1) & 0.3(2)\\
F2    & PG               &     4.0(4) &    3.4(4) & 0.3(2)\\
F2    & QF               &     1.0(1) &    1.2(1) & 0.2(2)\\
F2    & QG               &     1.0(1) &    1.1(1) & 0.2(2)\\
\hline
PMRA  & PF               &     3.6(4) &    1.4(1) & 0.3(2)\\
PMRA  & PG               &     3.6(4) &    1.3(1) & 0.2(2)\\
PMRA  & QF               &     1.0(1) &    1.1(1) & 0.1(2)\\
PMRA  & QG               &     1.0(1) &    1.1(1) & 0.1(2)\\
\hline
PMDEC & PF               &     4.3(5) &    1.0(1) & 0.0(2)\\
PMDEC & PG               &     3.6(4) &    1.1(1) & 0.0(2)\\
PMDEC & QF               &     1.0(1) &    0.9(1) & -0.1(2)\\
PMDEC & QG               &     1.0(1) &    0.9(1) & -0.1(2)\\

    \hline

    \end{tabular}

\end{table}

\subsection{Application to real data}
Although the simulations can demonstrate that the quasi-periodic model can improve results for a particular case, it is perhaps more informative to see the effect when applied to a sample of pulsars with known quasi-periodic variations. 
Therefore we apply the quasi-periodic model fitting to the sample of pulsars from \citet{lyne2010}, which exhibit strong quasi-period variations in their spin-down rate. These pulsars have recently been revisited by \citet{shaw22} (hereafter, S22) using more recent data, and this provides an ideal dataset with which we can test these models and investigate any effect on the pulsar parameters. We refer readers to S22 for details of the observations and preparation, but in brief the dataset consists of observations made with the 76-m Lovell telescope at Jodrell Bank, supplemented with observations from the 25-m `Mark-II' telescope also at Jodrell Bank. Most data prior to 2009 were centred at 1400-MHz and recorded using a 32-MHz analogue filterbank. Since 2009 most data are centred at 1520-MHz using a 384-MHz digital filterbank. These are supplemented with a small number of observations at 400, 610 and 925-MHz. For the digital filterbank data, radio frequency interference has been excised using a combination of median filtering and manual removal of affected channels or time intervals. ToAs are generated by cross correlation with a noise-free template using \textsc{psrchive}. A phase coherent timing solution is obtained using \textsc{tempo2}, making use of the pulse numbering feature to track the rotation of the pulsar over the entire dataset. Because of the inhomogeneity of backend instruments, we fit for separate white noise parameters (EFAC and EQUAD) for the legacy data, analogue filterbank data and digital filterbank data.

\begin{figure}
    \centering
    \includegraphics[width=\columnwidth]{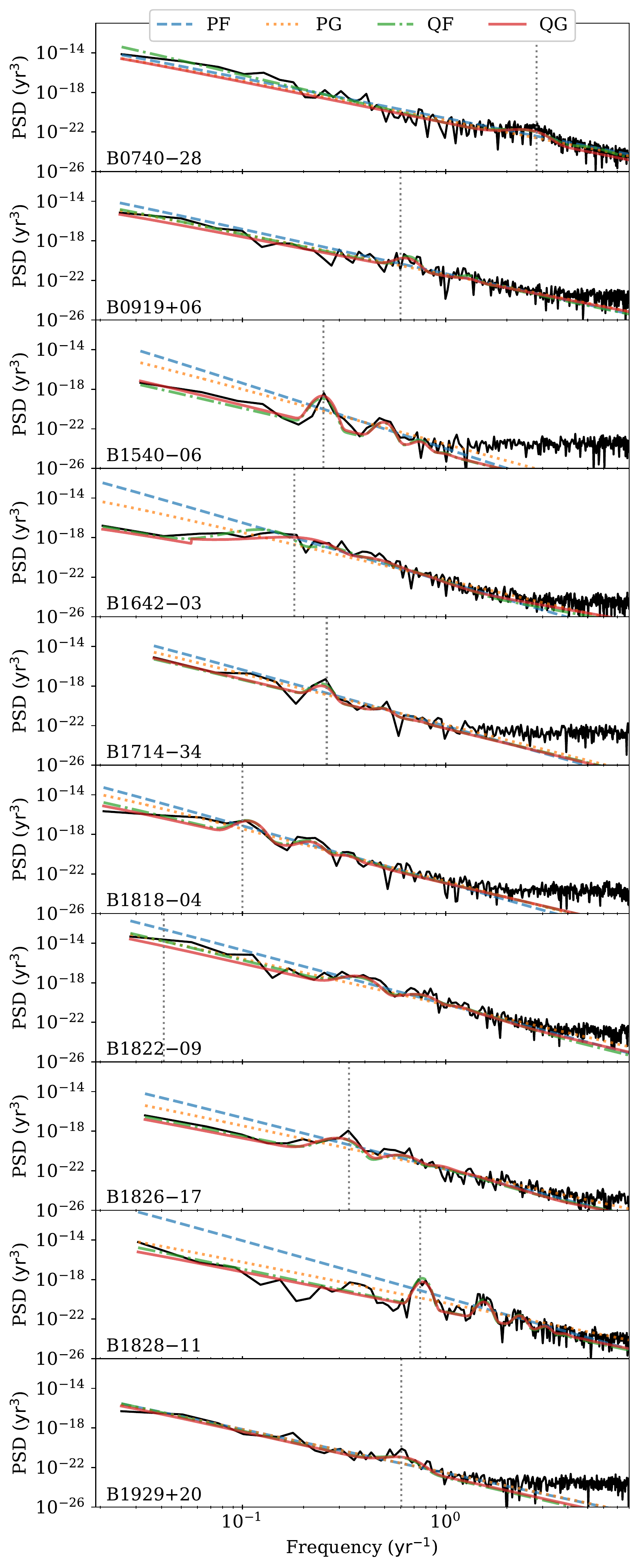}
    \caption{PSD of 10 pulsars estimated using the QG model overplotted with the model PSD from each of the four models. The dashed vertical line marks $f_\mathrm{qp}$.}
    \label{psdplot}
\end{figure}

We fit each of the models to the 17 pulsars in the dataset of S22. The parameter estimates for the QF quasi-periodic noise model are given in Table \ref{realpsr_period}. Results for the QG model are extremely similar and so are not repeated.
Ten of the pulsars show significant preference for the quasi-periodic model, with a log-Bayes factor, $\ln Z_\mathrm{QF} - \ln Z_\mathrm{PF}$, greater than 6.
Of these, nine find a periodicity that matches with the peak of the periodogram of $\dot{\nu}$ computed by S22.
The remaining pulsar is PSR B1822$-$09, for which we find a much shorter period of 1000 days compared to a peak of 9000 days in S22.
We attribute this difference to the large exponent of the power spectrum for the red noise in this pulsar.
The $\dot\nu$ time series used in S22 for the periodogram is the second derivative of the pulsar residual.
This process of taking derivatives will whiten the red noise \citep{coles}, and hence reveal the quasi-periodic oscillations.
However in the case where $\gamma$ is significantly larger than 4 there will continue to be residual red noise in the $\dot\nu$ timeseries, which likely leads to the identification of a periodicity consistent with the dataset length in S22 for PSR B1822$-$09.

\begin{table*}
    \centering
    \caption{Results of fitting the QF quasi-periodic model to a sample of pulsars. $\ln\left(Z_\mathrm{QF}/Z_\mathrm{PF}\right)$ is the log-Bayes factor in favour of a the quasi-periodic model. Columns 3 -- 8 give model parameters are as defined in Equations \ref{ppl_ent} and \ref{ppl_qp}.  $1/f_\mathrm{\dot{\nu}}$ is the period associated with the peak of the $\dot{\nu}$ periodogram in S22 for comparison. Values in parenthesis indicate the uncertainty in the last given digit. Pulsars are ordered by the evidence for the quasi-periodic model.}
    \label{realpsr_period}
    
    \begin{tabular}{lllllllllll}
   PSR & $\ln\left(\frac{Z_\mathrm{QF}}{Z_\mathrm{PF}}\right)$& $\log_{10}\left(\frac{f_\mathrm{qp}}{\mathrm{yr^{-1}}}\right)$ & $\log_{10}(R_\mathrm{qp})$ & $\lambda$ & $\sigma$ & $\log_{10}(A_\mathrm{pl})$ & $\gamma$ &  $1/f_\mathrm{qp}$ (day)& $1/f_\mathrm{\dot{\nu}}$ (day)& $f_\mathrm{\dot{\nu}}/f_\mathrm{qp}$\\
  
        \hline
B1828$-$11 &   $147.5$  &  $-0.111(2) $  &  $2.7(2)    $  &  $0.7(1)  $  &  $0.047(4)  $ &   $-9.41(3)    $  &  $4.3(1)$  &  $471(3)      $  &  $490(50)$  &  $0.97(9)     $\\
B0740$-$28 &   $25.5 $  &  $0.42(1)   $  &  $1.3(2)    $  &  $3(2)    $  &  $0.18(1)   $ &   $-9.37(4)    $  &  $4.9(2)$  &  $138(3)      $  &  $130(5)      $  &  $1.06(5)     $\\
B1540$-$06 &   $23.8 $  &  $-0.60(1)  $  &  $2.7(3)    $  &  $0.5(1)  $  &  $0.04(2)   $ &   $-11.2(3)    $  &  $5.1(7)$  &  $1470(30)$  &  $1461(6)     $  &  $1.01(2)     $\\
B1642$-$03 &   $18.7 $  &  $-0.82(5)  $  &  $2.6(3)    $  &  $1.6(3)  $  &  $0.17(2)   $ &   $-10.43(9)   $  &  $4.0(3)$  &  $2400(300)$  &  $2000(600)$  &  $1.2(4)      $\\
B1826$-$17 &   $16.7 $  &  $-0.50(2)  $  &  $1.9(3)    $  &  $2.0(5)  $  &  $0.14(3)   $ &   $-10.3(2)    $  &  $4.4(3)$  &  $1140(50)$  &  $1094(3)     $  &  $1.05(5)     $\\
B1822$-$09 &   $11.2 $  &  $-0.41(4)  $  &  $1.3(3)    $  &  $1.6(5)  $  &  $0.15(2)   $ &   $-9.4(2)     $  &  $5.5(4)$  &  $1000(100)$  &  $8900(200)$  &  $0.11(1)     $\\
B1929$+$20 &   $10.1 $  &  $-0.22(7)  $  &  $1.1(2)    $  &  $0.3(4)  $  &  $0.13(4)   $ &   $-10.3(1)    $  &  $4.7(2)$  &  $600(200)$  &  $604(3)      $  &  $1.0(3)      $\\
B0919$+$06 &   $8.8  $  &  $-0.19(2)  $  &  $0.9(2)    $  &  $2(1)    $  &  $0.12(4)   $ &   $-9.65(8)    $  &  $4.3(2)$  &  $570(50)$  &  $600(300)$  &  $0.9(5)      $\\
B1818$-$04 &   $6.5  $  &  $-0.97(2)  $  &  $1.7(3)    $  &  $0.7(2)  $  &  $0.06(4)   $ &   $-10.27(4)   $  &  $4.9(2)$  &  $3400(100)$  &  $4000(1000)$  &  $0.9(3)      $\\
B1714$-$34 &   $6.4  $  &  $-0.60(2)  $  &  $1.4(3)    $  &  $0.5(4)  $  &  $0.05(3)   $ &   $-10.0(1)    $  &  $5.2(4)$  &  $1460(60)$  &  $1400(10)$  &  $1.04(4)     $\\
B2148$+$63 &   $1.6  $  &  $-0.43(9)  $  &  $1.8(8)    $  &  $5(3)    $  &  $0.18(3)   $ &   $-11.6(5)    $  &  $4.8(7)$  &  $1000(200)$  &  $1300(200)$  &  $0.8(2)      $\\
B2035$+$36 &   $1.1  $  &  $-0.2(2)   $  &  $0.5(6)    $  &  $5(3)    $  &  $0.13(4)   $ &   $-9.5(1)     $  &  $5.9(3)$  &  $700(300)$  &  $13200(200)$  &  $0.05(3)     $\\
B1839$+$09 &   $0.8  $  &  $-0.2(3)   $  &  $0.3(9)    $  &  $5(3)    $  &  $0.11(6)   $ &   $-10.0(1)    $  &  $4.1(3)$  &  $800(800)$  &  $320(3) $  &  $2(2)        $\\
B1907$+$00 &   $0.0  $  &  $-0.7(3)   $  &  $0(1)      $  &  $3(3)    $  &  $0.10(6)   $ &   $-10.4(1)    $  &  $5.1(3)$  &  $2100(700)$  &  $5900(100)$  &  $0.4(1)      $\\
B0950$+$08 &   $-0.5 $  &  $-0.8(3)   $  &  $0(1)      $  &  $2(3)    $  &  $0.12(6)   $ &   $-11.3(1)    $  &  $5.5(3)$  &  $3000(1000)$  &  $5000(1000) $  &  $0.7(3)      $\\
B1903$+$07 &   $-1.0 $  &  $-0.5(3)   $  &  $-0.6(8)   $  &  $5(3)    $  &  $0.10(6)   $ &   $-9.06(7)    $  &  $4.6(2)$  &  $1500(800)$  &  $1800(900)$  &  $0.8(6)      $\\
J2043$+$2740 &   $-1.2 $  &  $-0.2(4)   $  &  $-0.9(7)   $  &  $5(3)    $  &  $0.10(5)   $ &   $-9.25(5)    $  &  $5.7(2)$  &  $800(600)$  &  $4010(900) $  &  $0.2(1)      $\\

        \hline

    \end{tabular}

\end{table*}

\begin{figure}
    \centering
    \includegraphics[width=\columnwidth]{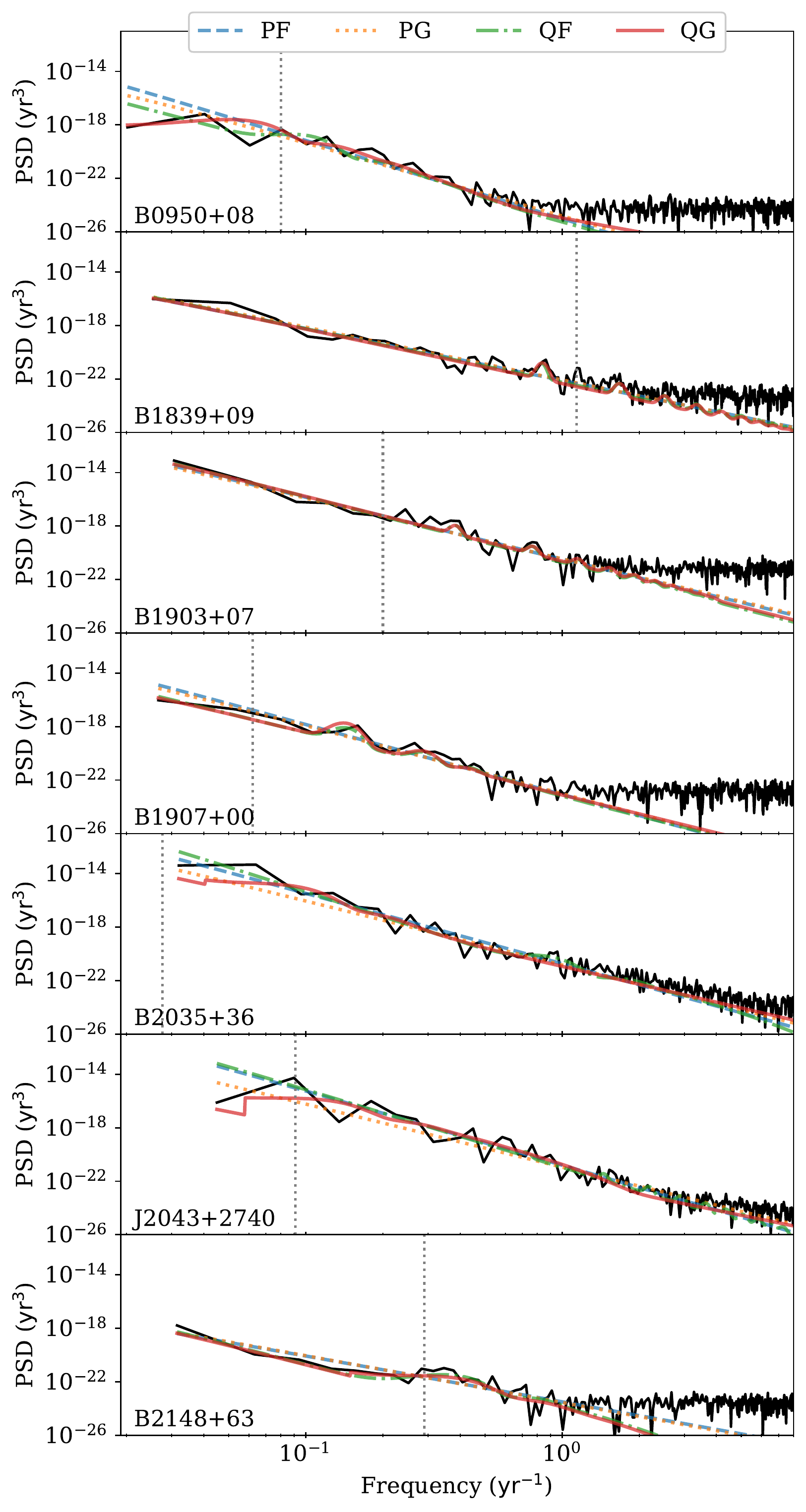}
    \caption{PSD of 7 pulsars without significant evidence for quasi-periodic variability, estimated using the QG model overploted with the model PSD from each of the four models.}
    \label{psdplot_noz}
\end{figure}

Figure \ref{psdplot} shows the PSD estimated from the residuals using the QG model as well as the maximum likelihood model PSD for each of the four models, for each of the pulsars with significant evidence in favour of a quasi-periodic model.
The remaining pulsars are shown in Figure \ref{psdplot_noz}.
For the pulsars in Figure \ref{psdplot}, the quasi-periodic models are able to capture the shape of the estimated PSD more closely than the pure power-law models, with the power-law models consistently over-estimating the power at the lowest frequencies.
The PF model typically has a greater over-estimation factor, in the most extreme case of B1828$-$11, over-estimating the PSD at $1/T_\mathrm{span}$ by nearly 4 orders of magnitude.
The post-fit F2, PMRA and PMDEC parameters for the 10 pulsars in Figure \ref{psdplot} are given in Table \ref{paramfit_table}.
As the main difference in the PSD models is at the lowest frequencies, the main impact is on the estimation of F2.
In the case of PSR B1828$-$11 using the QF model results in a $>10$-$\sigma$ measurement of F2, which is otherwise not significant under the PF model.
As we might expect, the proper motion measurements are most affected for pulsars with $f_\mathrm{qp}$ closest to 1 per year.
For example, PSR B1828$-$11 ($f_\mathrm{qp}\approx0.78\,\mathrm{yr}^{-1}$) sees a reduced uncertainty in proper motion parameters by a factor of around 4, whilst PSR B0740$-$28 ($f_\mathrm{qp}\approx2.8\,\mathrm{yr}^{-1}$) reduces the uncertainty in proper motion by a factor of 2.
These two pulsars also have the highest evidence in favour of the QF model.
PSR B1929$+$20 ($f_\mathrm{qp}\approx0.5\,\mathrm{yr}^{-1}$) sees a reduction in uncertainty of proper motion by a factor of 1.4.
The remaining pulsars show little change in the value or uncertainty for proper motion, but all have a much longer periodicity or do not have strong evidence in favour of the quasi-periodic model (hence the quasi-periodic variations do not dominate the timing noise).

\subsection{Spin-down variations}
The prevailing model for the quasi-periodic variations is that the spin-down rate of the pulsar varies in a quasi-periodic manner, hence it is useful to visualise the $\dot\nu(t)$ time-series from our quasi-periodic model.
The Fourier-domain Gaussian process model can be used to estimate $\dot\nu(t)$ by taking the second derivative of the Gaussian process model.
This can be computed analytically once we have the Gaussian process in the form
\[
r(t) = \sum\limits_i^{N_\mathrm{coef}} A_i \cos{\omega_i t} + B_i \sin{\omega_i t},
\]
where $A_i$, $B_i$ are the best-fit amplitudes of the Fourier-domain Gaussian process components from \textsc{tempo2} and $N_\mathrm{coef}$ is the number of Fourier coefficients.
The spin-down rate can therefore be computed analytically, since by our definition of residual we have
\[
\nu(t) =  \mathrm{F0} + \mathrm{F1} t+ \frac{1}{2}\mathrm{F2} t^2 -\mathrm{F0}\,\dot{r}(t),
\]
and hence,
\begin{equation}
\begin{split}
\label{nudotequation}
\dot\nu(t) &= \mathrm{F1}+\mathrm{F2} t-\mathrm{F0}\,\ddot{r}(t) \\
&=  \mathrm{F1} +\mathrm{F2} t + \mathrm{F0} \sum\limits_i^{N_\mathrm{coef}} A_i \omega_i^2\cos{\omega_i t} + B_i \omega_i^2 \sin{\omega_i t}.
\end{split}
\end{equation}
The uncertainty on $\dot\nu(t)$ can be estimated from the parameter covariance matrix, and the ToA uncertainties as outlined in Appendix \ref{appendixA}.

Figure \ref{nudotplot} shows the $\dot{\nu}(t)$ derived directly from the QF model in this work, compared against the results presented in S22 which are generated from a time-domain Gaussian process applied to the post-fit residuals.
There is generally a lot of similarity between the two methods for estimating $\dot{\nu}(t)$.
Some differences appear around glitches, e.g. in PSR B0740$-$28, B0919$+$06 and B1828$-$11.
The estimation of glitch parameters, particularly the change in $\dot{\nu}$ is difficult in the presence of quasi-periodic variations in $\dot{\nu}$ and hence the values obtained are sensitive to the choice of noise model used.
We claim that the $\dot{\nu}$ timeseries around the glitch in the two pulsars with the highest evidence for the QF model, PSR B0740$-$28 and B1828$-$11, look qualitatively more plausible for our QF model than that from S22, however PSR B0919$+$06 does not seem well modelled around the glitch for either case, perhaps reflecting other unmodeled transient glitch effects in this pulsar.

PSR B0950$+$08 shows a long-term deviation between our model and S22, a pulsar for which there is no evidence in favour of the QF model over the PF model.
The deviation from S22 is because the QF model behaves like a power-law with a low frequency turn-over for this pulsar, and hence deviates from the power-law significantly at the lowest frequencies (see Figure \ref{psdplot_noz}).
This causes a change in the estimation of F2 in this pulsar leading to a linear deviation in the estimated $\dot{\nu}(t)$, however neither measurement is significant. The maximum likelihood solution for F2 changes from $(1.4\pm1.1) \times 10^{-26}\mathrm{Hz^3}$ with the PF model to $(-0.4\pm1)\times10^{-26}\,\mathrm{Hz^3}$ with the QF model.
It is worth noting that this pulsar has variation timescale of several thousand days, and so only about two cycles are seen within our observing window.
It is possible that with a significantly longer observing span the evidence in favour of the quasi-periodic model would be more significant.

PSR B1822$-$09 and B2035+36 show rapid variations in the $\dot{\nu}(t)$ derived from our model compared to that in S22.
For these two pulsars the $\dot{\nu}$ timeseries do not appear to be well modelled by a Gaussian process, with evidence for non-stationarity in the statistics.
In these cases it appears the time-domain Gaussian process of S22 performs better than fitting directly to the residuals with the Fourier basis Gaussian process, though likely neither method really is optimal for this type of pulsar behaviour.

PSR B1818$-$04 shows high frequency oscillations in S22 that are not present in our model.
A periodicity analysis shows that these oscillations are consistent with a period of 1 year, which suggests an error in the position or proper motion in the timing model of S22.
Although S22 used a Gaussian process model of $\dot{\nu}(t)$, the timing model was solved in a `traditional' way, and is therefore susceptible to leakage from the spin noise into the pulsar parameters as described in \citet{coles}.
This highlights the benefits of the noise modeling procedure for determining pulsar parameters and studying the rotation.
A similar feature also appears in B1714$-$34, but to a lesser extent.

\begin{table*}
    \centering
        \caption{Post-fit timing parameters for a sample of pulsars when using the PF and QF models. The Bayes factor in favour of the QF model is also included comparison.}
    \label{paramfit_table}
    \begin{tabular}{lccccccc}
    \multirow{2}{*}{PSR} &\multirow{2}{*}{$\ln\left(\frac{Z_\mathrm{QF}}{Z_\mathrm{PF}}\right)$} & \multicolumn{2}{c}{F2 ($10^{-25}\mathrm{Hz}^3$)}  & \multicolumn{2}{c}{PMRA (mas/yr)} & \multicolumn{2}{c}{PMDEC (mas/yr)} \\
      & & PF & QF & PF & QF & PF & QF \\
         \hline
B1828$-$11 &   $147.5$  &  $12 \pm 21$  &  $11.3 \pm 0.8$  &  $-12 \pm 40$  &  $2 \pm 10$  &  $10 \pm 178$  &  $2 \pm 46$\\
B0740$-$28 &   $25.5$  &  $8.3 \pm 3.2$  &  $10 \pm 6$  &  $-28 \pm 15$  &  $-26 \pm 8$  &  $-16 \pm 20$  &  $-12 \pm 10$\\
B1540$-$06 &   $23.8$  &  $-1.2 \pm 2.4$  &  $0.04 \pm 0.07$  &  $-17.7 \pm 0.9$  &  $-17.4 \pm 0.8$  &  $-6.0 \pm 2.8$  &  $-4.2 \pm 2.4$\\
B1642$-$03 &   $18.7$  &  $0.3 \pm 1.2$  &  $-0.01 \pm 0.05$  &  $-2.6 \pm 1.3$  &  $-2.8 \pm 1.2$  &  $16 \pm 4$  &  $16 \pm 4$\\
B1826$-$17 &   $16.7$  &  $0 \pm 4$  &  $0.45 \pm 0.22$  &  $10 \pm 5$  &  $10 \pm 6$  &  $58 \pm 49$  &  $58 \pm 57$\\
B1822$-$09 &   $11.2$  &  $1 \pm 14$  &  $5 \pm 5$  &  $16 \pm 28$  &  $18 \pm 29$  &  $-100 \pm 110$  &  $-90 \pm 120$\\
B1929$+$20 &   $10.1$  &  $-0.45 \pm 0.33$  &  $-0.48 \pm 0.25$  &  $1.5 \pm 2.0$  &  $2.1 \pm 1.4$  &  $0.0 \pm 3.1$  &  $0.4 \pm 2.2$\\
B0919$+$06 &   $8.8$  &  $0.8 \pm 1.0$  &  $1.4 \pm 0.5$  &  $8 \pm 14$  &  $4 \pm 12$  &  $70 \pm 40$  &  $60 \pm 32$\\
B1818$-$04 &   $6.5$  &  $-0.2 \pm 1.1$  &  $0.09 \pm 0.18$  &  $-7.9 \pm 1.0$  &  $-8.0 \pm 0.9$  &  $15.5 \pm 2.9$  &  $15.6 \pm 2.9$\\
B1714$-$34 &   $6.4$  &  $-1 \pm 4$  &  $0.1 \pm 0.9$  &  $2 \pm 6$  &  $2 \pm 5$  &  $30 \pm 28$  &  $27 \pm 25$\\

\hline
    \end{tabular}

\end{table*}

\begin{figure*}
    \centering
    \includegraphics[width=2\columnwidth]{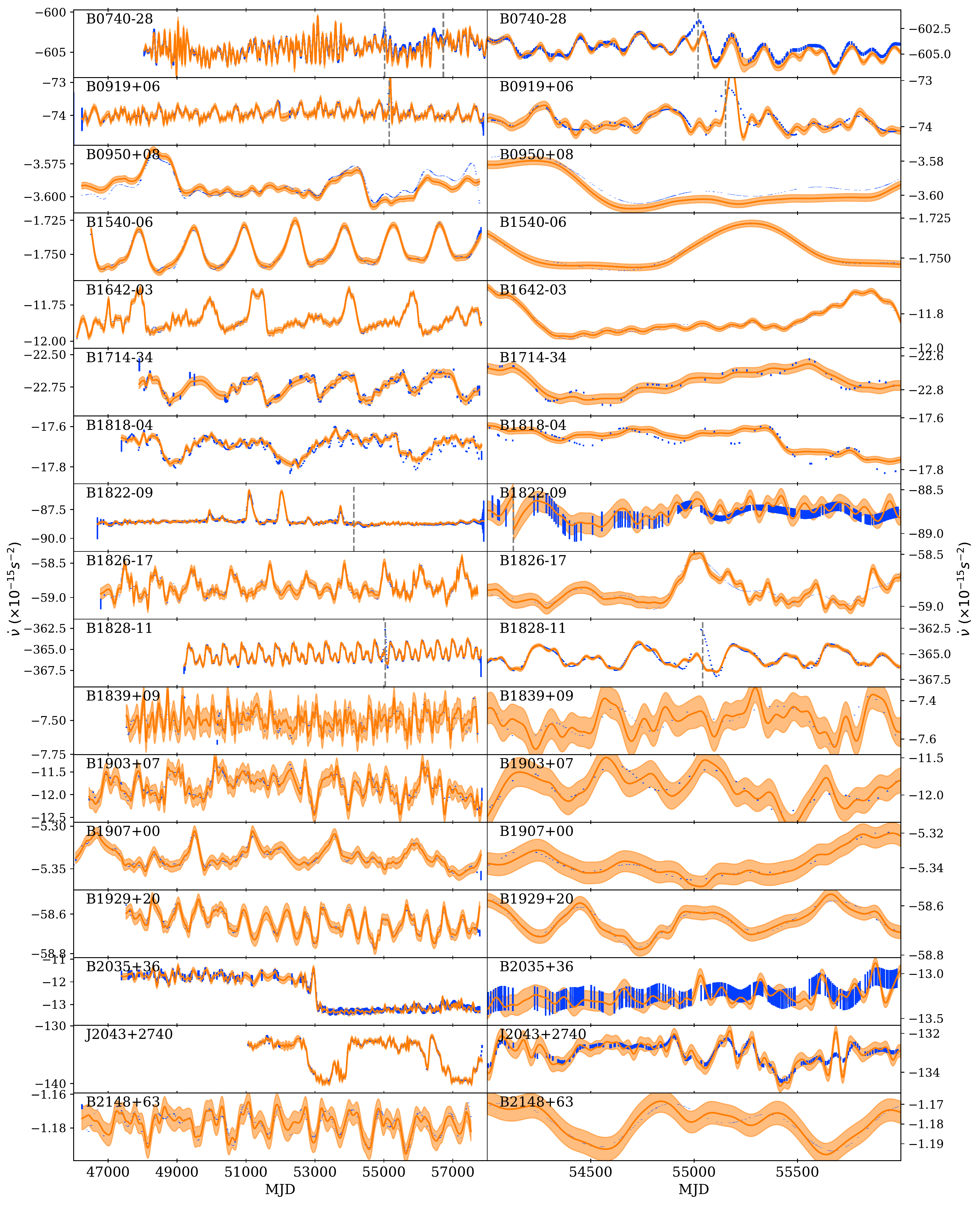}
    \caption{\label{nudotplot}Frequency derivative of the 17 pulsars from S22. Points with errors are those presented in S22 and the smooth curve is that derived from the QF model in this work, with the filled band representing the 1-$\sigma$ uncertainty. Vertical dashed lines mark the epoch of glitches. The left column shows the full range of the data and the right column shows a zoom between MJD 54000---56000 to highlight some of the details of the time variation.}
\end{figure*}

\section{Discussion}
As may be naively expected, the results of this work show that the parameter estimates are generally improved when the noise model more closely matches the underlying noise process.
However, the widely used power-law Fourier domain Gaussian process models are seen to be very robust and it takes fairly significant deviations to cause issues.
Particularly the quasi-periodic model is only significantly advantageous in a small number of pulsars, even when we study a sample of pulsars known to exhibit significant quasi-periodic variations.
Hence we expect that it is not necessary to apply a quasi-periodic model to most pulsars, and it will be clear from inspection of the $\dot{\nu}$ time-series or power-spectra computed from a power-law model if the quasi-periodic model is likely to be of any benefit to the study of a given pulsar.

Based on the simulations in Section \ref{sec:lowfreq}, we see that the Fourier basis model with low-frequency cut-off at $1/T_\mathrm{span}$ is not suitable for estimating F2 for pulsars with steep timing noise. Indeed many pulsars do show timing noise with a power-law exponent at or above $\gamma=4$ (e.g. \citealp{asj+19}), hence some mitigation is needed for this.
Here we propose that a simple adjustment of the low-frequency cut-off to $1/2T_\mathrm{span}$ significantly improves the results at large values of $\gamma$, though this does increase computation as twice as many frequency coefficients are needed to reach the same value of $f_\mathrm{high}$.

Although this work focuses on the estimation of F2 and other pulsar parameters for the study of pulsars, this may also have implications for the ongoing gravitational wave pulsar timing array experiments that are searching for 
a steep spectrum signal in the pulsar data.
Here we have shown that quasi-periodic variations in the pulsar spin down can lead to a spurious over-estimation of the modelled power at low frequencies.
Although the actual spectral estimates are not typically over-estimated, except perhaps in the most extreme cases, this would falsely increase the uncertainty from pulsar spin noise and potentially the total power estimated in the gravitational wave background (GWB).
Of course, the pulsars forming the pulsar timing array do not typically show evidence for quasi-periodic oscillations in spin down, so this effect is unlikely to be significant, but it does demonstrate that the choice of model is important.
Perhaps more importantly, the choice of how the model should behave at frequencies below $1/T_\mathrm{span}$ clearly does have significant impact on the parameter estimates and it is unlikely that the pulsar timing noise really does cease or turn-over at frequencies around $1/T_\mathrm{span}$.
This is especially important for pulsars exhibiting steep red noise.
We can use the estimation of F2 as a proxy for our ability to model the lowest frequencies in the data, since an error in the F2 estimate is caused by unmodelled low frequency red noise.
We find that the critical power-law exponent for reliable measurements of F2 with modeled $f_\mathrm{low}=1/T_\mathrm{span}$ is around $\gamma=4$, which is also the exponent that would be produced by random variations in $\dot{\nu}$ as well as that expected from the GWB.
This is consistent with the break in the uncertainty of measuring F2 against red noise power-law exponent observed by \citet{xkbs19}.
Therefore we argue that the choice of $f_\mathrm{low}$ can be important for the gravitational wave experiment, particularly as these analyses typically do not include F2 in the fitting.
This means that any unmodelled power in the pulsar spin noise may be falsely attributed to another source, perhaps manifesting in the requirement of additional model terms, such as the recently observed common uncorrelated red noise process \citep{ipta_dr2_gw}. A full exploration of the impact of over-sampling the spectrum for GWB searches is outside the scope of this paper, but there may be other benefits to oversampling the Fourier space by capturing second order statistics in the data, and we suggest further exploration of these effects should be included in confirming the robustness of a future GWB detection.

\section{Conclusions}
We set out to answer two questions regarding the choice of noise model for pulsar timing.
For the case of the choice of low-frequency cut-off and the periodic boundary conditions, we found that below about $\gamma = 4$ all the models perform well even with the lowest frequency set to $1/T_\mathrm{span}$.
However, many pulsars show significantly steeper red processes, and hence the power must either be absorbed into F2 or another mitigation must be applied if F2 is a parameter of interest.
Here we propose a simple solution of setting the lowest frequency as $1/2T_\mathrm{span}$, which performs reliably up to about $\gamma=6$, beyond which the model again struggles.

For quasi-periodic timing variations, we find that a small number of pulsars are better modelled with our new noise model that includes a quasi-periodic term.
However, the power-law models generally do well even without this quasi-periodic term for many of the pulsars, and it is only for the pulsars with the strongest quasi-periodic variations that there is a significant issue with measuring F2 with the power-law model.
We find proper motion measurements can be affected where the quasi-periodic fluctuations have a period of the order of a year.
We believe it is unlikely that quasi-periodic models will significantly improve pulsar timing if the quasi-periodic signal is not already clearly apparent in the power spectrum of the residuals.

We also confirm previous results that in general, the Fourier-basis Gaussian process performs largely as well as the equivalent GLS model.
The new models and implementation of the GLS method are all made available for testing and real-world use through \textsc{run\_enterprise}, which will allow for use, or further testing, on a wide range of data.

\section*{Acknowledgements}
Pulsar research at Jodrell Bank is supported by a consolidated grant from the UK Science and Technology Facilities Council (STFC).
ICN is also supported by the STFC doctoral training grant ST/T506291/1. 
The authors thank Ben Shaw for providing the comparative data from S22.
The authors thank the reviewer for providing insightful comments at peer review.
\section*{Data Availability}

The raw data in this work are as used in \citet{shaw22}, and requests to use these data should be made to the original authors of that work.
The $\dot{\nu}$ timeseries from this work (i.e. those shown in Figure \ref{nudotplot}) are available at \href{http://dx.doi.org/10.5281/zenodo.7664166}{doi:10.5281/zenodo.7664166}.



\bibliographystyle{mnras}
\bibliography{mtsp} 




\appendix

\section{Implementation of the Gaussian process model in tempo2 and estimation of the uncertainties}
\label{appendixA}
The \textsc{tempo2} software performs the final step of evaluating the timing model and is used in this work to extract the Gaussian process time-series and the spin-down timeseries.
\textsc{Tempo2} uses generalised least squares fitting, though for the Gaussian process model the data covariance matrix, $\mathbfss{N}$, is assumed to be diagonal and contain the white noise uncertainties for each data point on the diagonal.
A design matrix $\M$ maps the parameters $\boldsymbol{\beta}$ to a model of the residuals, $\boldsymbol{r}$ observed at at times $\boldsymbol{t}$.
Following the standard least-squares methodology, we could compute an estimate of the parameters from the residuals,
\[
\boldsymbol{\hat\beta} = (\M^\tr\mathbfss{N}^{-1}\M)^{-1}\M \mathbfss{N}^{-1}\boldsymbol{r}.
\]
In practice this is performed using either the singular-value decomposition (SVD), or the QR decomposition to avoid the numerical precision issues with direct computation.
Similarly, we can compute the covariance matrix of the parameter estimates,
\[
\C_{\beta} = (\M^\tr\mathbfss{N}^{-1}\M)^{-1},
\]
which again is in practice computed via the SVD or QR decomposition.

To fit the Fourier domain Gaussian process with $n$ coefficients to $m$ data points, $\M$ will contain $m$ rows and $2n$ columns with the form
\[
M_{j,i} = \begin{cases}
        \sin{(2\pi f_i t_j)} \text{\,for $0 < i < n$,}
        \\
        \cos{(2\pi f_{i-n} t_j)} \text{\,for $n < i < 2n$}.
        \end{cases}
\]

In order to implement the Gaussian priors on these parameters, we add $2n$ constraints to the least-squares problem.
This is achieved by extending $\boldsymbol{r}$ with $2n$ zero valued elements, and extending $\M$ with $2n$ block diagonal elements that contain the Gaussian likelihood

\[
M_{i+m,i+2n} = \begin{cases}
        P(f_i)^{-1/2} \, \delta\!f\text{\,for $0 < i < n$},
        \\
        P(f_{i-n})^{-1/2} \,\delta\!f \text{\,for $n < i < 2n$},
        \end{cases}
\]
where $P(f_i)$ is the model PSD at frequency $f_i$ and $\delta\!f$ is the frequency resolution of the Fourier components, which scales from PSD to power. This constrained problem can then be solved using the least squares formalism.
In practice $\M$ will also contain columns associated with the pulsar timing model (e.g. the standard spin and astrometric parameters), however for the remainder of this discussion it is assumed that we will extract only the relevant rows and columns related to the red noise model.
The linear algebra from this point forward is implemented in the \textsc{make\_pulsar\_plots.py} script provided with \textsc{run\_enterprise}.

Estimates of the $\dot\nu(t)$ at each epoch can be made by computing $\D$, the second time derivative of $\M$ scaled by $\mathrm{F0}$ (cf. Equation \ref{nudotequation}),
\[
D_{j,i} = \begin{cases}
        4\pi^2 f_i^2 \mathrm{F0}\,\sin{(2\pi f_i t_j)} \text{\,for $i < n$,}
        \\
        4\pi^2 f_{i-n}^2 \mathrm{F0}\,\cos{(2\pi f_{i-n} t_j)} \text{\,for $n < i < 2n$},
        \end{cases}
\]

In the same way, we can also compute a matrix $\D_m$ that maps the parameters onto an arbitrary vector of time samples $\boldsymbol{t}_m$, allowing us to interpolate the value of $\dot\nu(t)$ at any arbitrary time by

\[
\dot\nu(\mathbf{t}_m) =  \D_\mathrm{m} \hat{\boldsymbol{\beta}}.
\]

In order to compute the uncertainty on $\dot\nu(t)$ we first need the covariance matrix of Gaussian process in the absence of any observations,
\[
\C_\mathrm{mm} =  \D_\mathrm{m} \C_\beta \D_\mathrm{m}^\tr.
\]
We also need the covariance between the $\dot\nu(t)$ and the observations,
\[
\C_\mathrm{mo} = \D_\mathrm{m} \C_\beta \M^\tr,
\]
and the covariance of the data with itself
\[
\C_\mathrm{oo} = \M \C_\beta \M^\tr + \mathbfss{N},
\]
These can then be combined to estimate the covariance of the model $\dot\nu(t)$ in the presence of the observations,
\[
\C_{\dot\nu} = \C_\mathrm{mm} - \C_\mathrm{mo} \C_\mathrm{oo}^{-1} \C_\mathrm{mo}^\tr.
\]
We take the diagonal elements of $\C_{\dot\nu}$ to draw representative uncertainties on the estimates of $\dot\nu(t)$.
Note that this does not include any uncertainty related to the uncertainty on the model hyperparameters, but this can be estimated by repeating the process for a range of samples from the hyperparameter posteriors.

\bsp	
\label{lastpage}
\end{document}